%% file: acl_latex.tex
\pdfoutput=1

\documentclass[11pt]{article}

\usepackage{acl}

\usepackage{times}
\usepackage{latexsym}

\usepackage[T1]{fontenc}

\usepackage[utf8]{inputenc}

\usepackage{microtype}

\usepackage{inconsolata}

\usepackage{graphicx}
\usepackage{stfloats}

\usepackage[bottom]{footmisc}

\title{PyTorch-IE\\ \vspace*{.5\baselineskip} \textbf{Fast and Reproducible Prototyping for Information Extraction}}

\author{
    Arne Binder \and Leonhard Hennig \\ 
    German Research Center for Artificial Intelligence (DFKI) \\
    \{\textit{arne.binder, leonhard.hennig}\}\textit{@dfki.de}
    \AND
    Christoph Alt \\ 
    \textit{christoph.alt@posteo.de}
}

\begin{document}
\maketitle


\input{sections/00_abstract}

\input{sections/01_introduction}
\input{sections/02_design_goals}
\input{sections/03_architecture_new}
\input{sections/04_details_new}
\input{sections/05_predefined_modules_new}

\input{sections/06_conclusion_new}

\bibliography{custom_and_anthology_selected}

\end{document}

%% file: sections/00_abstract.tex
\begin{abstract} 

The objective of Information Extraction (IE)
is to derive structured representations from unstructured or semi-structured documents. 
However, developing IE models is complex due to the need of integrating several subtasks.
Additionally, representation of data among varied tasks and transforming datasets into task-specific model inputs presents further challenges. To streamline this undertaking for researchers, we introduce PyTorch-IE, a deep-learning-based framework uniquely designed to enable swift, reproducible, and reusable implementations of IE models. PyTorch-IE offers a flexible data model capable of creating complex data structures by integrating interdependent layers of annotations derived from various data types, like plain text or semi-structured text, and even images. We propose task modules to decouple the concerns of data representation and model-specific representations, thereby fostering greater flexibility and reusability of code. PyTorch-IE also extends support for widely used libraries such as PyTorch-Lightning for training, HuggingFace datasets for dataset reading, and Hydra for experiment configuration. Supplementary libraries and GitHub templates for the easy setup of new projects are also provided. 
By ensuring functionality and versatility, PyTorch-IE provides vital support to the research community engaged in Information Extraction.

\end{abstract}

%% file: sections/01_introduction.tex
\section{Introduction}

\begin{figure}[htb!]
  \includegraphics[width=0.49\textwidth,keepaspectratio,trim={0cm 2cm 10cm 0cm},clip]{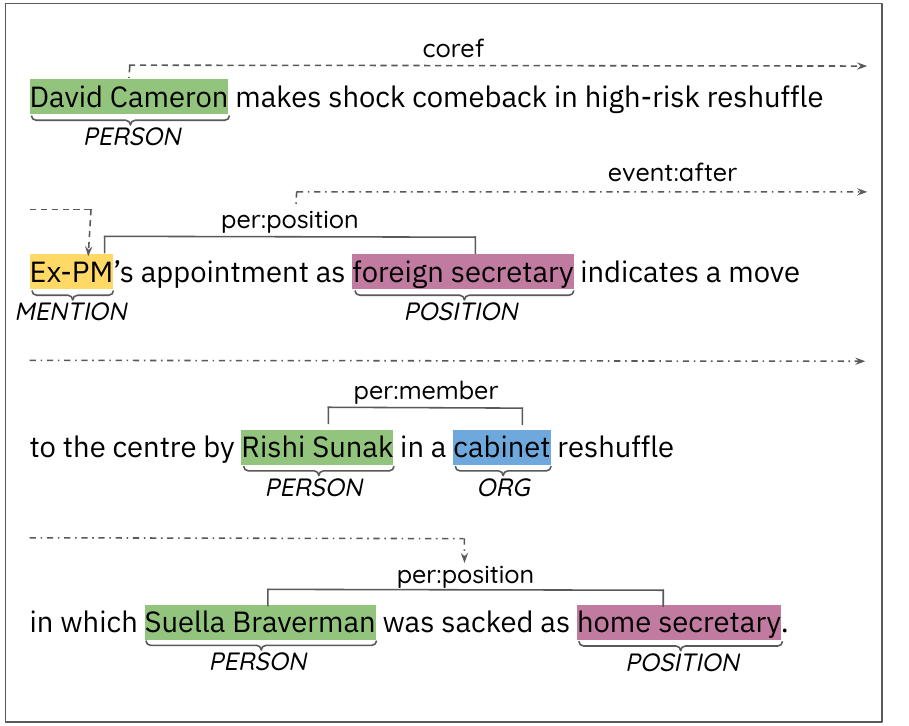}
  \caption{
  In Information Extraction, structured representations of data are frequently organized using a data schema that encapsulates layered annotations, where higher-layer annotations connect annotations from lower layers. However, how data is organized differs from how it is represented as input for task-specific model architectures, necessitating a need for translating between data schema and model inputs/outputs (see \S \ref{sec:framework}).
  }
  \label{fig:example_layered_annotations}
\end{figure}

Information Extraction (IE) covers a wide range of tasks that aim to extract structured representations from unstructured or semi-structured documents. In order to construct composite, higher-order representations, IE approaches often combine several sub-tasks, e.g., Named Entity Recognition (NER), Entity Linking (EL), and Relation Extraction (RE) for Knowledge Base Population~\cite{dong-etal-2014-knowledge,thomas-etal-2017-streaming,trisedya-etal-2019-neural,Sui2020JointEA,qin-etal-2021-erica,ren-etal-2021-hyspa,de-cao-etal-2022-multilingual}. Extracted information is typically represented and shared between sub-tasks by some form of data schema that structures interdependent layers of task-specific \emph{annotations}, e.g., token- and span-level labels for entities or part-of-speech tags, and structural links between basic elements for e.g., dependency parses or event extraction~\cite{thomas-etal-2017-streaming,schon-etal-2018-corpus,du-etal-2021-template,lu-etal-2021-text2event}. Figure~\ref{fig:example_layered_annotations} shows an example with annotations for entities, relations, and event relations.

Many recent IE approaches rely on popular libraries such as PyTorch~\cite{paszke-etal-2019-pytorch} and HuggingFace Transformers \cite{wolf-etal-2020-transformers}, as well as platforms such as the HuggingFace Hub\footnote{\href{https://huggingface.co}{huggingface.co}}, to train, evaluate and share models and datasets. However, these libraries focus on general purpose (PyTorch) and Transformer-based (HuggingFace) deep learning, with relatively little support for complex, joint, and pipeline-oriented IE tasks and information sharing via data schemas. Implementations of IE approaches hence often needlessly re-invent the wheel by defining their own data schemas and implementing task-specific, non-reusable solutions for dataset loading and conversion to model inputs.

In this work, we introduce PyTorch-IE\footnote{\href{https://github.com/ChristophAlt/pytorch-ie}{github.com/ArneBinder/pytorch-ie}}, a Python framework that supports fast development of deep-learning-based Information Extraction approaches. Following its design goals (\S \ref{sec:design_goals}), PyTorch-IE greatly simplifies the development of reproducible, reusable, and extensible IE approaches. It complements existing machine learning libraries with a flexible \emph{data model} that enables the representation of IE-centric complex data structures, realized as interdependent annotation layers of the underlying source text. However, the data model is not limited to a particular set of IE sub-tasks or specific model architectures. To translate between data schema and task-specific encoded model representations, PyTorch-IE also introduces the concept of \emph{task modules}. Task modules convert, and thus decouple, data representation from the task-specific model representations, enabling greater flexibility and re-usability of data preparation and model code (\S \ref{sec:framework}). 
We leverage a simple use-case to demonstrate how these concepts translate into re-usable modules (\S \ref{sec:use_case}).

\begin{figure*}[!t]
  \includegraphics[width=\textwidth]{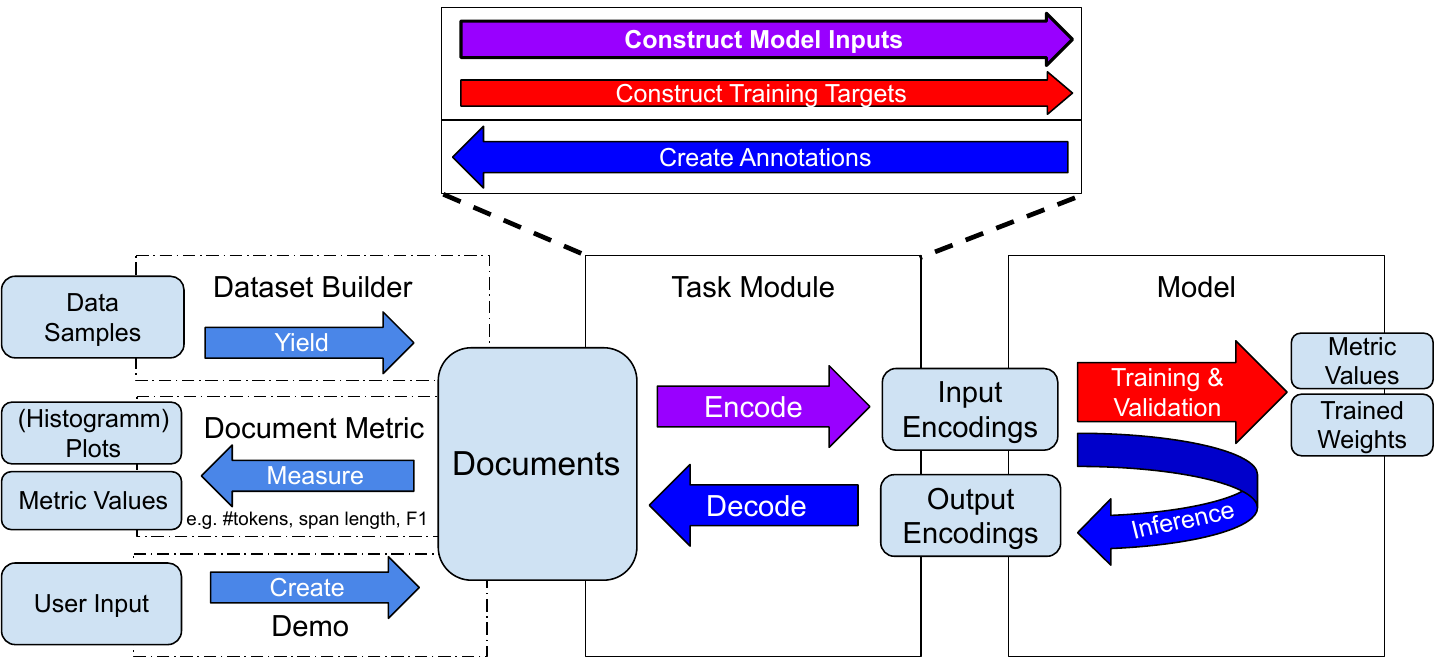}
  \caption{PyTorch-IE Architecture: Main components and data flow. At its core, the \textbf{Task Module} transforms documents to model inputs and training targets (encode) and integrates model outputs back in form of new annotations (decode). \textbf{Documents} follow a user-defined, typed data schema. They form an interface for rapid dataset integration, analysis of prediction quality as well as dataset statistics via document metrics, and ease the creation of demo applications. \textbf{Models} are arbitrary Lightning modules. }
  \label{fig:core_architecture}
\end{figure*}

PyTorch-IE is not restricted to plain text, but supports Information Extraction from semi-structured text (e.g., HTML), two-dimensional text (e.g., OCR'd images), and images. 
It uses external well-maintained libraries for non-core functionality: PyTorch-Lightning\footnote{\href{https://lightning.ai}{lightning.ai}} for training and logging, HuggingFace datasets~\cite{lhoest-etal-2021-datasets} for dataset reading, and Hydra~\cite{Yadan2019Hydra} for experiment configuration. All models developed in PyTorch-IE can be easily shared via the HuggingFace model hub, which also allows to quickly build demos\footnote{see here for examples: \href{https://huggingface.co/pie}{huggingface.co/pie}
} based e.g., on HuggingFace spaces, Gradio, or Streamlit.

In addition to the core framework, the PyTorch-IE ecosystem  includes additional libraries with implementations of models, task modules, and metrics as well as dataset builders that work out-of-the-box for common IE tasks (e.g. Named Entity Recognition, Relation Extraction, Extractive Question Answering). It also includes a GitHub template based on the Lightning-Hydra-Template\footnote{\href{https://github.com/ashleve/lightning-hydra-template}{github.com/ashleve/lightning-hydra-template}} to bootstrap deep-learning-based Information Extraction projects (\S \ref{sec:family}).





%% file: sections/02_design_goals.tex
\section{Design Goals} \label{sec:design_goals}

In this section, we present the design goals that underpin our framework. 

\paragraph{Reproducibility} Our project is driven by a research-oriented mindset. We aim to ensure that it is possible to revisit and execute the training process even after several years, yielding the same results.

\paragraph{Reusability} We believe in avoiding reinventing the wheel for every new research idea. By providing modules that work out-of-the-box and contribute to the research landscape, we not only share experimental results and insights but also enable others to readily utilize our work.

\paragraph{Extensibility} Our project aims to support the easy reproduction of existing approaches while also providing a streamlined process for efficiently implementing new ideas.

\paragraph{Leveraging Existing Frameworks} We strive to build upon well-known frameworks, focusing on areas that have not been adequately addressed yet.

%% file: sections/03_architecture_new.tex
\section{Framework Architecture} \label{sec:framework}

The PyTorch-IE framework is designed with a strong focus on reproducibility and reusability. The structure of the framework allows a clear separation of model code from data preprocessing code, meaning that implementation of models and handling of complex data can happen independently. As IE data typically has complex semantics - for instance, events defined over entity spans with distinct attributes - the framework is built around a flexible data model that can represent various data structures with different annotation layers.

Figure~\ref{fig:core_architecture} gives an overview of the architecture and data flow of PyTorch-IE. At its core, data is represented as typed documents which store base data and its semantic annotations (\S~\ref{subsec:data_schema}). Task modules (\S~\ref{subsec:task_module}) encode documents for a task-specific model (\S~\ref{subsec:training}) and decode its output back into annotations. Furthermore, PyTorch-IE eases the development of metrics (\S \ref{subsec:metrics}) to measure prediction performance or to gather dataset statistics. In addition, it comes with a set of well established usability features (\S~\ref{subsec:usability}) attached. Figure~\ref{fig:core_architecture} presents an overview of the PyTorch-IE components and how they connect to each other.

\subsection{Document Model} \label{subsec:data_schema} PyTorch-IE provides a straightforward, yet flexible data schema. Central to this is the concept of a \textit{document type}, which defines the base data for the data points and the specific annotations assignable to them.

The schema is designed to efficiently manage the structural intricacy of IE data through typed base data and typed annotations. This offers the benefit of early identification of any structural complexity-related errors by static type checking, thereby preventing extensive debugging.

A document can encompass one or numerous base \textit{data fields} such as text or image data. The annotations are 
organised into \textit{annotation layers}, which reference a single or several data fields as well as other annotation layers. These layers are, therefore, capable of forming arbitrary non-cyclic annotation graph structures.
Annotation layers explicitly state their references in the layer definition. This configuration serves two purposes: 1) ensuring correct (de-)serialization of the documents, achieved by using the resultant annotation graph to determine the sequence of object instantiation, and 2) allowing dependent annotations to access referenced base data (or annotations) directly from within the dependent annotations. 

Each annotation layer incorporates a dedicated field for model predictions. This keeps gold and predicted values together and, therefore, simplifies metric calculation and error analysis.

\subsection{Task Modules} \label{subsec:task_module} 
We introduce the notion of a \textit{task module}, designed to encapsulate the functionality required for converting documents into model inputs, and reconciling model outputs back into their originating source documents.
This proposition is inspired by the observation that documents, characterized by identical semantic structures, can serve as training sources for various models such as sequence classification, token classification, and token generation, among others. 
More critically, adhering to the same document type and model while modifying the encoding methods continues to gain importance, especially with the advent of pre-trained, context-dependent embedding models like BERT \cite{devlin-etal-2019-bert}.

This concept's significance becomes even more pronounced in the realm of Information Extraction (IE) as determining the optimal approach to encoding structural information often entails ambiguity. Different encoding strategies that exemplify this point include task-label text pairs for few-shot classification \cite{halder-etal-2020-task}, text that incorporates entity marker tokens for RE \cite{baldini-soares-etal-2019-matching}, and the use of pointer and label triples for output in generative named entity recognition \cite{yan-etal-2021-unified-generative}.

In the context of textual input, a task module may specify
tokenization, long input handling, and the integration of particular markers to encode span data or other features.
Additionally, working with datasets composed of multiple, semantically different but structurally similar information---such as coreference detection and RE or claim identification and NER data---requires the selection and encoding of data considered relevant based on the specific method employed.

In detail, PyTorch-IE provides concepts to structure the \textit{encoding}, i.e., the conversion of documents to model inputs, and the \textit{decoding}, i.e., the integration of model outputs back into the respective source documents. Especially the encoding is done in two steps: 1) (model) \textit{input} encoding and 2) (training) \textit{target} encoding. The former holds the core logic of the task module and specifies how to construct the model inputs from a document for a certain task by using the base data and its annotations. For instance, it defines how to combine the textual context with the entity information so that a relation classification model can perform its prediction on it. The latter, the encoding of the target data that will be used to fit or test the model, is executed during model training and evaluation, but omitted during inference. This leads to an inherent separation of concerns, preventing training information leakage.

\subsection{Trainable Models} \label{subsec:training} For effective parameter optimization and evaluation, we utilize the Lightning framework, which significantly reduces boilerplate code compared to using PyTorch directly. 
Any Lightning module can function as a model in PyTorch-IE. It requires just a few additional lines of code to let it seamlessly work with the model serialization and sharing functionality provided through the Huggingface Hub.

\subsection{Metrics}\label{subsec:metrics}  Our standardized data schema facilitates the creation of reusable metric implementations. With PyTorch-IE, annotations are comparable by design; they are frozen dataclass objects, with metadata such as probability scores excluded from comparison. Thus, it is possible to derive sets of true positives, false negatives, and false positives directly using Python set operations.

\subsection{Usability Features} \label{subsec:usability} 

\paragraph{Sharing Models} Trained models can easily be shared on the HuggingFace Hub\footnote{\href{https://huggingface.co/}{huggingface.co}}. Uploading a trained model as well as a task module configuration is simply enabled by setting a flag. PyTorch-IE's auto-model and auto-taskmodule features allow for one-line instantiation of trained models and pre-configured task modules shared on the HuggingFace Hub or from a local directory.

\paragraph{HuggingFace Datasets Integration} Loading data is made convenient through the integration with HuggingFace's Datasets\footnote{\href{https://github.com/huggingface/datasets}{github.com/huggingface/datasets}}, which supports various functionalities like caching of pre-processing results and streaming of large datasets.

\paragraph{Logging Frameworks} The framework supports various logging platforms such as Weights \& Biases\footnote{\href{https://wandb.ai/}{wandb.ai}}, Aim Stack\footnote{\href{https://aimstack.io/}{aimstack.io}}, and Tensorboard\footnote{\href{https://www.tensorflow.org/tensorboard}{www.tensorflow.org/tensorboard}}. Integration is made possible through a configuration-based setup.

\paragraph{Code Quality} PyTorch-IE is designed to yield high-quality, reliable code. An easily extendable test suite ensures that the provided code works as expected. Code quality checks are performed via pre-commit\footnote{\href{https://pre-commit.com}{pre-commit.com}} hooks, ensuring that any additions or changes maintain the standards of the existing codebase.

%% file: sections/04_details_new.tex
\section{Use Case: Building a Relation Extraction model} \label{sec:use_case}

In this section, we briefly present a 
use case illustrating how to leverage various components of PyTorch-IE to handle a specific task: binary relation extraction by classifying text that is annotated with entity mentions.

\subsection{Document Design} 
As explained in \S \ref{subsec:data_schema}, a document consists of data fields and annotation layers. We specify a single field \texttt{text} for the base data. We create two annotation layers, \texttt{entities} and \texttt{relations}. The \texttt{entities} layer consists of labeled spans that reference the \texttt{text}, with the attributes \texttt{start}, \texttt{end}, and \texttt{label}. The \texttt{relations} layer holds binary relation annotations that refer to the \texttt{entities} layer, where each entry requires \texttt{head}, \texttt{tail}, and \texttt{label} attributes.

\begin{figure}[ht!]
  \includegraphics[width=0.49\textwidth,keepaspectratio,clip]{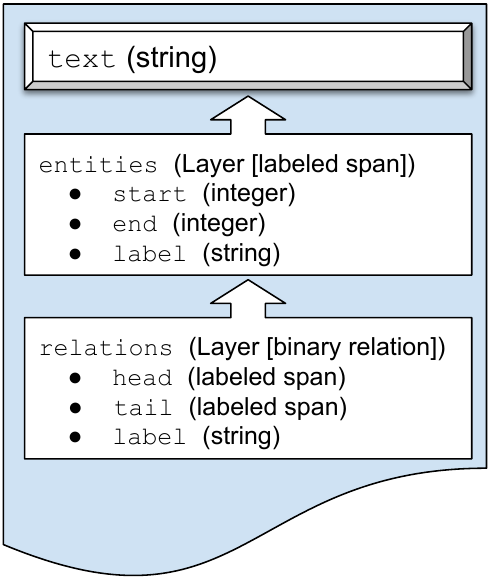}
  \caption{
  Simple document schema for relation extraction data consisting of a single data field (\texttt{text}) and two annotation layers (\texttt{entities}, \texttt{relations}). Types are shown in brackets, annotation attributes as list items and arrows indicate layer dependence. Note that PyTorch-IE supports more sophisticated structures (multiple data fields and arbitrary, non-cyclic annotation graphs).  
  }
  \label{fig:example_document_schema}
\end{figure}

As described above, it is vital to specify the dependencies between annotation layers. In our scenario, \texttt{entities} relies on \texttt{text}, while \texttt{relations} depends on \texttt{entities}. Thus, an entity annotation can leverage its reference to render a string representation that exhibits the relevant text content instead of only displaying its start and end indices. Figure~\ref{fig:example_document_schema} presents the final document schema for the example task.   

\subsection{Task Module}

In PyTorch-IE, the task module's primary role revolves around transforming documents into recognizable model inputs. Assuming a binary relation \texttt{rel} with entities \texttt{head} and \texttt{tail} as its arguments, the following steps outline a valid approach for our relation extraction task:

\begin{enumerate} 
    \item Tokenize the text. 
    \item Insert special marker tokens before and after both entities \texttt{head} and \texttt{tail}. These tokens depend on the argument type, which could be either \texttt{HEAD} or \texttt{TAIL}, and may also hinge on the entity type, e.g., \texttt{ORG} or \texttt{PER}. 
\end{enumerate} 
Subsequently, feed the token sequence to any sequence classification model, e.g., a Transformer model from HuggingFace. This process can be tailored in multiple ways to adapt to varying encoding approaches.

During the input encoding phase, we can also implement additional logic to handle possible complexities like: 
\begin{itemize} 
    \item Managing instances where the text length exceeds the maximum input size of the model. Depending on specific requirements or constraints, strategies could range from applying a sliding window to using a fixed context centered around the relation arguments. 
    \item Intelligently generating candidate relations by selecting a subset of all possible entity pairs. 
\end{itemize}

\subsection{Model Definition} \label{subsec:model_details} 
Lastly, we develop a model tailored to our use case. This would typically involve devising a simple Lightning module encapsulating a Transformer-based \cite{NIPS2017_3f5ee243} sequence classification model from HuggingFace Transformers \cite{wolf-etal-2020-transformers}. 
Leveraging their auto-model functionality enables us to use a wide range of pre-trained models that are available on the HuggingFace Hub by merely exchanging the model identifier.

%% file: sections/05_predefined_modules_new.tex
\section{The PyTorch-IE Ecosystem} \label{sec:family}

The PyTorch-IE environment is an ecosystem encompassing several essential projects designed to facilitate research projects in Information Extraction. These projects include the PyTorch-IE core framework along with the basic models, task modules, and metrics. Additionally, the PyTorch-IE ecosystem extends to PyTorch-IE-Models\footnote{\href{https://github.com/ArneBinder/pie-modules}{github.com/ArneBinder/pie-modules}} and PyTorch-IE-Datasets\footnote{\href{https://github.com/ArneBinder/pie-datasets}{github.com/ArneBinder/pie-datasets}} that offer additional models, task modules, metrics, dataset builders, and document conversion utilities. Finally, the PyTorch-IE-Hydra-Template\footnote{\href{https://github.com/ArneBinder/pytorch-ie-hydra-template-1}{github.com/ArneBinder/pytorch-ie-hydra-template-1}} functions as a GitHub template, which streamlines the creation of model training and evaluation pipelines and offers a modular configuration system to kick-start new research ventures.

\subsection{Ready-Made Solutions} The PyTorch-IE environment equips users with pre-configured model architectures and preprocessing procedures tailored for tasks such as Named Entity Recognition, Relation Extraction, and Extractive Question Answering. Users can utilize the pre-trained models available in the HuggingFace's Model Hub as a starting point and subsequently train them on a variety of datasets.

\paragraph{Models and Task Modules:} We provide models and task modules for a range of tasks, including 
NER (token classification, span classification), RE (sequence classification with entity augmentation), and joint modeling tasks (simple sequence-to-sequence (seq2seq), seq2seq with pointer networks). 
They can be easily adapted to other span detection and classification tasks. For further reference and potential implementations, users are encouraged to visit the PyTorch-IE-models repository.

\paragraph{Datasets:} PyTorch-IE also provides pre-processed datasets for tasks like NER (for instance, Conll2003, Conll2012), RE (for instance, Tacred), and EQA (like SQuAD v2.0), along with generic ones (like BRAT\footnote{\href{https://brat.nlplab.org/standoff.html}{brat.nlplab.org/standoff.html}}). More dataset options and a guide how to integrate your own data are available at PyTorch-IE-datasets and PyTorch-IE's HuggingFace page.\footnote{\href{https://huggingface.co/pie}{huggingface.co/pie}}

\paragraph{Metrics:} In addition to metrics for model output evaluation such as F1-Score, PyTorch-IE encompasses diverse metrics for statistical dataset analysis. These include, for instance, metrics for assessing label distribution and input length distribution in terms of characters or tokens. For more model and dataset metric implementations, you can refer to PyTorch-IE-models and PyTorch-IE-datasets.

%% file: sections/06_conclusion_new.tex
\section{Conclusion}

This paper introduces PyTorch-IE, a versatile and adaptable framework designed to assist the rapid yet robust development of Information Extraction approaches that utilize deep learning. This solution offers a high degree of reproducibility, builds upon well-established ML libraries, and adventures into hitherto unexplored areas of such tasks. PyTorch-IE fosters extensibility and reusability, simplifying the process of replicating existing implementations and crafting new strategies.

At its core, PyTorch-IE carefully separates model implementation and high-level data preprocessing, fostering flexibility and adaptability. It presents a flexible data schema and task module system that supports intricate IE tasks while providing the capacity to accommodate future tasks within this area of study, including those connected with data structures not yet conceptualized.

Additionally, PyTorch-IE bolsters usability by facilitating seamless model sharing, easy loading of different datasets, support for a plethora of logging frameworks, and a commitment to high-quality, reliable code. It comes equipped with modules offering out-of-the-box functionality and provides a range of models, task modules, and metrics implementation for various common IE tasks.

In conclusion, PyTorch-IE marks a stride in the bid to simplify the development of reproducible, reusable, and extensible Information Extraction approaches. It creates a bridge between complex data structures inherent to IE tasks and powerful pre-trained language models, pushing forward the boundaries in the field of automated IE research. The modular architecture of PyTorch-IE motivates the exchange of ideas and methods in the research community and opens up new avenues for collaborative development.